\documentstyle[11pt,aasms]{article}
\lefthead{??}
\righthead{}

\def\lta{{\>\rlap{\raise2pt\hbox{$<$}}\lower3pt\hbox{$\sim$}\>}}
\def\gta{{\>\rlap{\raise2pt\hbox{$>$}}\lower3pt\hbox{$\sim$}\>}}

\lefthead{??}
\righthead{}
 
\begin{document}

\title {Spiral galaxies with {\tt WFPC2}: II. The nuclear properties of 40 objects \footnote{Based on observations with the NASA/ESA Hubble
Space Telescope, obtained at the Space Telescope Science Institute,
which is operated by Association of Universities for Research in
Astronomy, Inc.\ (AURA), under NASA contract NAS5-26555}}

\author{C.M.\ Carollo\footnote {Hubble Fellow}$^,$\footnote{Johns
Hopkins University, 3701 San Martin Dr., Baltimore, MD 21218
}$^,$\footnote{Space Telescope Science Institute, 3700 San Martin Dr.,
Baltimore, MD 21218}, M. Stiavelli$^{4,}$\footnote {On assignment
from the Space Science Dept. of the European Space
Agency}$^,$\footnote {On leave from the Scuola Normale Superiore,
Piazza dei Cavalieri 7, I56126 Pisa, Italy} \&
J. Mack$^4$ }

\begin{abstract}
We report the analysis of HST {\tt WFPC2} F606W images of 40 spiral
galaxies belonging to the sample introduced in Carollo et al. (paper
I), where 35 other targets were discussed. We describe the optical
morphological properties of the new 40 galaxies, derive the surface
brightness profiles for 25 of them, and present the results of
photometric decompositions of these profiles into a ``bulge''
($R^{1/4}$ or exponential) and a disk component.  The analysis of the
enlarged sample of 75 galaxies puts on a statistically more solid
ground the main results presented in paper I, namely: {\it I.}  In
$\approx$30\% of the galaxies, the inner, morphologically-distinct,
structures have an irregular appearance. Some of these ``irregular
bulges'' are likely to be currently star forming. {\it II.}  Resolved,
central compact sources are detected in about 50\% of the galaxies.
{\it III.}  The central compact sources in galaxies with nuclear star
formation are brighter, for similar sizes, than those in non star
forming galaxies. {\it IV.}  The luminosity of the compact sources
correlates with the total galactic luminosity.  \\ Furthermore, the
analysis of the enlarged sample of 75 objects shows that: {\it A.}
Several of the non-classical inner structures are well fitted by an
exponential profile. These ``exponential bulges'' are fainter than
$R^{1/4}$ bulges, for given total galaxy luminosity and (catalog)
Hubble type later than Sab.  {\it B.}  Irregular/exponential bulges
typically host central compact sources.  {\it C.} The central sources
are present in all types of disk galaxies, starting with systems as
early as S0a. About 60\% of Sb to Sc galaxies host a central compact
source. Many of the galaxies which host compact sources contain a
barred structure.  {\it D.}  Galaxies with apparent nuclear star
formation, which also host the brightest compact sources, are
preferentially the early- and intermediate-type (S0a-Sb) systems. {\it
E.} None of the features depends on environment: isolated and not
isolated galaxies show indistinguishable properties. \\ Independently
from the physical nature of the non-classical inner structures, our
main conclusion is that a significant fraction of galaxies classified
from the ground as relatively early-type spirals shows a rich variety
of central properties, and little or no morphological/photometric
evidence for a smooth, $R^{1/4}$-law bulge.

\end{abstract}


\section{Introduction}

This is the second of a series of papers dedicated to the
investigation of the (nuclear) properties of spiral galaxies. To this
aim, we have started a HST {\tt WFPC2} snapshot survey in the F606W
filter of a sample of 92 spiral galaxies (complemented by 15
additional early-type galaxies, for a comparison with the literature),
75 of which have already been observed. In the first paper of the
series (Carollo et al.\ 1997a, paper I), we have discussed the optical
nuclear morphology (and surface brightness profiles) for 35 of our
target galaxies, i.e. the ones observed until the HST servicing
mission of February 11, 1997. In this paper, we do the same for the
additional 40 targets which have been observed since then, the last of
the observations being dated April 15, 1997. Analogously to what done
in paper I, we use our measurements to investigate the occurrence of
nuclear spiral structure, the properties of the bulge-like components
in the inner regions of disk galaxies, the frequency of occurrence of
nuclear compact sources, and the relation of the above properties with
global galactic properties. Here, the enlarged sample of 75 galaxies
allows us to perform a statistical analysis of the frequency of
occurrence of the various features as a function of luminosity, Hubble
type and environment.

In contrast to the wealth of information on the nuclei of early-type
galaxies provided by HST (Crane et al.\ 1993; Jaffe et al.\ 1994;
Lauer et al.\ 1995; Carollo et al.\ 1997b, 1997c; Faber et al.\ 1997),
at the start of our investigation little was known either about the
nuclear properties of early-type disk galaxies, or about the relation
between the nuclear properties and those of the surrounding galactic
structure, e.g., the bulge and/or the inner disk (see the recent
review by Wyse, Gilmore \& Franx 1997). Phillips et al.\ (1996)
presented WF/PC images for a sample of 20 spirals. That analysis was
hampered by the inferior angular resolution of the pre-{\tt COSTAR}
HST; furthermore, the Phillips et al.\ sample was strongly biased
toward late-type galaxies. The exploration of the innermost regions of
early-type spirals, i.e. of spheroidals at the low-end side of the
luminosity sequence, promises to uncover important aspects regarding
the formation of nuclei, and their dependence on environment and
dynamical structure.  More generally, it may provide a crucial test
for bulge -- and galaxy -- formation scenarios.

This paper is organized as follows. In Section 2 we briefly review the
criteria that we adopted to choose the sample of galaxies, and the
steps of data reduction performed to derive the surface brightness
profiles. In Section 3 we present the morphological properties of the
40 new galaxies, the analytical fits used to derive the luminosities
and sizes of their bulges, and the procedures adopted to quantify the
luminosities and sizes of their nuclear compact sources. In contrast
to paper I, in which the bulge-like components were fitted exclusively
with a $R^{1/4}$-law, here we fit the inner bulge-like structures with
either a $R^{1/4}$-law or an exponential profile, and report the
parameters for the best fitting analytical description. In order to
ensure homogeneity in the quantification of the bulge properties, we
re-performed the analytical fits also for the galaxies presented in
paper I (see Appendix A).  In Section 4 we use the enlarged sample of
75 objects to discuss the nuclear versus global properties in
early-type spiral galaxies. We summarize our conclusions in Section 5.

\section{Sample Description and Data Analysis}

A sample of 92 spirals (plus 15 objects chosen among E/S0 and S0/Sa,
so as to have a reference sample for comparison with the literature on
early-type galaxies) was selected from two catalogs with well-defined
diameter limits (the UGC catalog for the northern hemisphere, Nilson
1973, and the ESOLV catalog for the southern hemisphere, Lauberts \&
Valentijn 1989). The objects were chosen to have: {\it (i)} Angular
diameter larger than 1 arcmin; {\it (ii)} Regular types {\it Sa, Sab,
Sb,} and {\it Sbc}; {\it (iii)} Reliable classification; {\it (iv)}
Redshift $<$ 2500 km/s; {\it (v)} An inclination angle, estimated from
the apparent axial ratio, smaller than 75 degrees. From the resulting
sample, we excluded the galaxies already imaged with HST, to optimize
the overall investment of HST time.  Seventy-five out of the total 107
targets have been observed to date. Of these, 35 were observed from
March 1996 up to the February 11th, 1997 HST servicing mission, and
constitute the sample for the study presented in paper I. Forty more
galaxies were observed from then on, and are presented in this
paper. Their apparent B magnitude, distance (derived from the RC3 --
de Vaucouleurs et al.\ 1991 -- redshift unless a mean group redshift
was available), environment (for those objects that are not isolated),
IRAS flux, full-width HI lines measured at the 20\% level, and
UGC/ESOLV and RC3 morphological classification, are listed in Table 1.
We adopt in the following analysis the Hubble types obtained from the
RC3 classification by using the provided conversion table (Table 2 of
RC3, which gives $0=S0a$, $1=Sa$, $2=Sab$, $\ldots$, $9=Sm$), since
the a posteriori inspection of the {\tt WFPC2} images shows that these
give generally a better description of the galaxies than the one given
by the UGC/ESOLV types.  Consistently with paper I, we here adopt
$H_0=65$ km/s/Mpc.

For each galaxy we acquired two {\tt WFPC2} F606W exposures of 400 and
200 seconds, respectively, with the galaxy nucleus centered on the PC.
Surface photometry was carried out in IRAF by using the {\tt STSDAS}
task ``ellipse''. Isophotal fits were restricted to the PC chip.
Details on the isophotal fitting procedure and on the technique used
to correct for patchy dust absorption are extensively described in
paper I. We were able to derive reliable surface brightness profiles
for 25 galaxies. The ubiquitous presence of bright knots in the
nuclear regions and strong dust obscuration did not allow us to derive
an acceptable isophotal fit for the remaining 15 galaxies.
 
\section{Results}

Figure 1 displays the inner $18'' \times 18''$ of the F606W images for
the new 40 galaxies.  For four of the objects, we show in Figure 2 the
unsharp-masking images, which better display their relevant nuclear
features.  The surface brightness profiles for the 25 galaxies for
which we were able to perform the isophotal fits are shown in Figure
3.  The substructure visible in most profiles is due, in some cases,
to a residual influence of the dust absorption or the star forming
regions on the derived profiles; in others, it indicates distinct
galactic components identifiable in the images.

\subsection{The Inner Structure}

The 40 new galaxies show a large variety of nuclear morphologies,
similarly to the 35 objects we presented in paper I.  Strong knot-like
enhancements of luminosity (almost certainly due to star formation)
are observed on spiral arms in the nuclear regions of several
galaxies.  In some galaxies the nuclear star formation occurs in round
``rings" or in $m=1$ spiral arms similar to the star forming
rings/arms of nuclear starbursts identified by e.g., Maoz et al.\
(1996).  In several objects spiral structure, as traced by spiral-like
dust lanes or the bright knots, reaches down to the innermost
accessible scales. In the entire sample of 75 objects, only a fraction
($\approx 40\%$) of the galaxies contains a smooth, classical bulge.
The central structure of several other objects has instead an
irregular morphology: it is often dwarf-irregular like; in a few cases
it is rather elongated, similar to a late-type bar. Occasionally,
small ``bulge-like'' features coexist with nuclear spiral
structure. We name these non-classical inner structures ``irregular
bulges'', within the caveats of a morphological classification, which
does not allow us to distinguish whether these features are actually
``bulges'', or rather bars, or even morphologically-distinct parts of
the disks.  The irregular bulges often host very compact, resolved
central sources. In several cases, they are full of bright knots,
suggesting that nuclear star formation might also be concentrated in
these structures.  In $\approx 30\%$ of the sample it is not clear
whether a (regular/irregular) bulge is present.  According to this
rich variety of structure, and consistently with our analysis in paper
I, we have grouped the new 40 galaxies in several categories,
depending on their nuclear morphology, i.e.  we have classified
galaxies as hosts of: {\it (i)} regular bulges (RB); {\it (ii)}
irregular bulges (IB); {\it (iii)} resolved central compact sources
(CS); {\it (iv)} unresolved, i.e. point-like, nuclear sources (PS);
{\it (v)} nuclear spiral structure (NSS); {\it (vi)} concentrated
nuclear star formation (NSF). The categories are listed by their
abbreviated nomenclature in Table 2.  Question marks indicate
ambiguous identifications due to e.g., strong dust obscuration, star
formation, inclination effects. The column ``Bar'', lists whether a
(early- or late-type, nuclear or large-scale) bar is detected (``Y'')
or not detected (``N'') in the {\tt WFPC2} images; a question mark
indicates that the presence of a bar is uncertain (e.g., in several
galaxies, it is hinted by a square morphology). In a few cases the
``bar'' identifies the same inner, elongated structure listed as an
``irregular bulge''.  As we stressed in paper I, it is important to
note that at the resolution of e.g., the Palomar Observatory Sky
Survey, the images of Figure 1 would have only a few elements, and
they would display a much more limited variety of morphologies.

\newpage
\subsection{Analytical Fits to the Bulge-Like Components}

In order to estimate quantitatively the physical scale and luminosity
of the bulge-like components present in the inner galactic regions,
and in contrast with paper I (where we adopted exclusively an
$R^{1/4}$-law representation for the inner galactic structure), we
fitted the PC surface brightness profiles with {\it (i)} an
$R^{1/4}$-law plus an exponential disk, {\it (ii)} the sum of two
exponential profiles, {\it (iii)} a single-exponential and {\it (iv)}
a single $R^{1/4}$-law profile. For those galaxies hosting a central
compact source, i.e. a distinct nuclear component ($\lta 0.3''$) in
the surface brightness profile, this was excluded from the fits.  The
theoretical laws were convolved with the HST Point Spread Function
(PSF), derived with Tinytim (Krist 1992) before performing the
fits. The use of simulated PSFs obtained by construction at the
nominal focus position is not necessarily inferior to the use of
archival stars, given that pointlike sources with adequate S/N located
near to the nuclei were not available for most of the galaxies, and
that focus drifts and breathing modify the PSF profile and affect the
flux within a 1 PC pixel radius up to 10\% (and within 5 PC pixels up
to 5\%; Suchkov \& Casertano 1997). Our approach of convolving the
models rather than deconvolving the data helps to minimize the effects
of using a possibly non-perfect PSF.

For each galaxy, the best of the above fits was adopted as descriptive
of its surface brightness profile.  In Table 2 we report in the column
``Fit" the half-light radius $R_e$ in arcseconds and the total
magnitude of the inner, i.e. ``bulge'', components identified by the
fits.  For the exponential bulges, $R_e \simeq 1.678 h$, with $h$ the
exponential scale length. The bulge parameters for the galaxies of
paper I used in the present analysis are given in Table 3 (see
Appendix A).  For the two-component fits, the uncertainty on the best
fit bulge parameters contains a contribution due to possible
``leaking'' of flux into/from the disk. We carried out a number of
tests in order to assess the robustness of the best fit flux
assignment to the bulge. In particular, we performed for each galaxy
several bulge-disk decompositions by varying the radial range of the
fits, or by fitting the bulge component for a range of predefined disk
parameters. The corresponding bulge magnitudes vary by up to 0.3
magnitudes, and the effective radii up to $\sim$30\%; in many cases
the uncertainty is smaller than these values.  These errors do not
affect our conclusions.  In Tables 2 and 3, the superscripts
$R^{1/4}$, $R^{1/4}$-{\it expo}, {\it expo1} and {\it expo2} close to
the galaxy names identify galaxies whose PC light distribution has
been fitted with an $R^{1/4}$-law, an $R^{1/4}$-law plus exponential
profile, a single- or a double-exponential profile, respectively.  The
analytical best fits to the surface brightness profiles are shown in
Figure 3 (solid lines for $R^{1/4}$ bulges, and dashed lines for
exponential bulges).

The galaxies which are fitted by a single exponential profile have
central F606W surface brightness in the range $\approx 19$-20 mag
arcsec$^{-2}$, i.e., on the bright-end side but compatible with
Freeman's law (1970) for disks. In spite of this, in eight out of ten
cases, there is morphological evidence that the single-exponential
fits describe only the inner structure of systems which likely have
two distinct components.  The morphology of the inner structure often
resembles that of dwarf elliptical galaxies, although somewhat more
irregular. A few of these systems have a late-type classification in
the RC3; however, similarly to their earlier-type relatives, they
clearly display in our {\tt WFPC2} images a morphologically distinct
central structure filling most of the PC chip, embedded in a typically
quiescent, i.e., non star forming, disk or halo. The latter is clearly
visible in the WF chips: it has low surface brightness (typical of the
disks described by Bothun et al.\ 1997), and shows little (or
sometimes no) spiral arm structure.  The incompleteness of the
isophotes, and the low signal-to-noise ratio of these faint outer
components in our relatively short exposures, make it impractical to
derive for them reliable isophotal fits.  Nonetheless, the
two-component appearance of these galaxies supports the association of
the ``dominant'' inner structures identified by the single-exponential
fits with ``exponential bulges''.  The case of E499G37, a galaxy
belonging to the sample of paper I, is ambiguous. This galaxy,
similarly to the other eight, hosts a PC-scale dominant
single-exponential bulge, which is embedded in a surrounding faint
structure. However, it also hosts the most extended central compact
source (comparable in size with the smallest exponential bulges;
Appendix A; Table 3). For this object, we retain the
single-exponential fit to describe its ``bulge'' and the fit derived
in paper I to the inner structure to describe its ``compact source''.
The only galaxy fitted by a single-exponential profile which shows no
evidence for a distinct outer component is ESO482G17.  For it, we
report in Table 3 both the single-exponential best fit parameters
describing the entire system, and the bulge parameters derived from a
double-exponential fit performed to obtain an upper limit to a
possible inner ``bulge'' component, for which however there is no
morphological evidence.

In contrast with the ``dominant, single-exponential bulges'', the
inner components of the double-exponentials fits are small, faint
photometrically-distinct structures embedded in dominant, typically
spiral-armed and star forming, disks.  In Figure 4 we plot the {\tt
WFPC2} mosaiced images for four representative galaxies, hosts of the
two classes of ``exponential bulges''. The figure illustrates the
difference between the disk-dominated galaxies fitted by
double-exponentials (lower panels: NGC4030, left; NGC6384, right), and
the ``bulge-dominated'' objects fitted by single-exponential profiles
(upper panels: ESO508G34, left; NGC4980, right).  It is possible that
relatively faint ``exponential bulges'' could be hidden also inside
the bright $R^{1/4}$-law bulges. For example, the addition of an
exponential component similar to the bulges of e.g., ESO498G5, NGC1325
and NGC2082 would not significantly affect the surface brightness
profile of a galaxy like NGC2196, and therefore would not be detected
in our photometric study. By contrast, relatively bright exponential
bulges such as those of NGC3177, NGC4030 and NGC6384 would affect the
inner light profile of NGC2196, which would significantly deviate from
an $R^{1/4}$-law.

\subsection{The Resolved Central Compact Sources}

In order to quantify the luminosities of the compact sources 
seen in the centers of several galaxies, we used three different
approaches. The first two approaches were already used and described
paper I, and are here summarized: (1) the contribution from the
compact source was fitted with a Gaussian, assuming that the
underlying galactic light continuum is well represented by the
asymptotic value of the Gaussian wings; and (2) the underlying
galactic light was modeled with the task ``fit/flat" in {\tt
ESO/MIDAS}, this model was subtracted from the image, and the counts
in excess were attributed to the central source.  In addition to these
two sets of measurements, we adopted a third criterion for performing
such measurements, namely (3) an iteratively smoothed and compact
source masked version of the original frame was subtracted from the
latter, and the measurement was performed on the resulting image
(containing only the compact source).  The three methods are
complementary: they sample the galactic background in different
regions, and hence their combined use allows us to estimate the
uncertainty in the derived values contributed by the poorly
constrained underlying galactic light.  We adopted as final the
averages of, and as errorbars the mean of the differences between,
these three estimates for the luminosity of the central sources.  The
sizes of the compact sources were taken equal to the FWHM of the best
fitting Gaussians of method (1), corrected for the instrumental width
of {\tt WFPC2}.  The so-derived sizes and luminosities are reported in
Table 2 (column ``Compact Source").  The third method was also applied
to the galaxies presented in paper I, and confirmed the estimates for
the compact sources parameters presented there. This ensures the
homogeneity of the measurements within the enlarged sample used in our
discussion.

In several (often $R^{1/4}$-law) galaxies, e.g., NGC2196, NGC3054,
NGC3277, NGC5121, NGC7421, NGC7690, the presence of a compact source
is ambiguous, since their central peak in density could be interpreted
either as a compact source, or as the inward extrapolation of the
outer light profile.  For the cases in which the ambiguity is less
pronounced and the images show a more convincing evidence for a
distinct central structure, e.g., NGC3277, NGC7421 and NGC7690, we
list their compact source parameters in Table 2 and append a question
mark to the ``CS'' entry in the last column.  For the remaining, less
convincing cases, we determined the brightest compact source that
could be subtracted from the galaxy without leaving a ``hole'' in its
light distribution. These upper limits are listed in Tables 2 and 3
(column ``Compact Source'').

\subsection{Additional Information on the Individual Galaxies}

In addition to the information coded in Table 2, we briefly report
below further comments relative to the (nuclear) properties of some of
the target galaxies. For the bulge-dominated galaxies whose PC light
profile has been described by a single exponential, we report also the
visible size of the faint, outer disk-like structure.

\smallskip

\noindent
{\it ESO~205G7.} A small bulge-like component is embedded in the
inner bar, and hosts in its very center a multi-armed spiral (radius
$\sim0.5''$; star forming?). The nucleus of the inner spiral is
possibly unresolved. Dust lanes along the bar connect to the central
spiral.

\noindent
{\it ESO~240G12.} The outer, faint and irregular component extends to
about six exponential scales (i.e., $\approx35''$) of the fitted
single-exponential bulge.

\noindent
{\it ESO~317G20.} Spiral-like dust lane reach close to the very
center. Tightly-wound (ring-like) star forming spiral arms are present
outside $\sim4''$ from the center (i.e., inside the fitted $R_e$).
Unclear whether the $R^{1/4}$-law structure is a real bulge or an
$R^{1/4}$-law disk (see Kormendy 1993 for an extensive discussion on
``dense disks'' in the centers of spiral galaxies).

\noindent
{\it ESO~443G80.} Star formation is concentrated in nuclear,
i.e. bulge-like, region.

\noindent
{\it ESO~508G34.}  The distinct, boxy single-exponential bulge is
rounder than the surrounding faint disk. The latter extends to roughly
six exponential scales (i.e., $\approx35''$) of the bulge, and shows
hints for weak spiral structure.

\noindent
{\it ESO~549G18.} The fitted exponential bulge is embedded in
the present bar-like structure.

\noindent
{\it ESO~572G22.}  No obvious spiral arms are detected in the faint,
quiescent outer disk, which extends to roughly eight exponential
scales (i.e., $\approx50''$) of the single-exponential bulge.  The
latter has a very elongated, bar-like morphology. Star formation is
concentrated in the nuclear region.

\noindent
{\it IC~879.} Highly elongated, star forming structure embedded in the
bar-like component. The faint ``spiral arm'' may be a tidal feature
(NGC 5078 is 2.5 arcmin away).

\noindent
{\it IC~1555.} Irregular, boxy inner morphology. Dust lanes cut the very
center.  

\noindent
{\it IC~5256.} Highly elongated, star forming structure embedded in
the bar-like component.

\noindent
{\it NGC~406.} Clear spiral arm structure extending to about ten
exponential scales ($\approx75''$) of the fitted single-exponential
bulge. The bulge might be star forming.

\noindent
{\it NGC~1325.} Spiral-like dust structure down to the nucleus.

\noindent
{\it NGC~1353.} Dust lanes down to the nucleus. The presence of a very
small ($\lta 1''$) bulge, partly obscured by dust lanes, cannot be
excluded.

\noindent
{\it NGC~1385.}  Elongated, star forming structure embedded in the
(irregular) bar-like component.

\noindent
{\it NGC~1640.}  Spiral-like dust lane down to the nucleus.  The
bulge-like structure is embedded in an early-type bar, and is fitted
by the sum of an $R^{1/4}$-law (inner component, reported in Table 2)
and an exponential (outer component, equivalent $R_e \sim 12''$). The
two components are morphologically indistinguishable.

\noindent
{\it NGC~2082.} Irregular spiral-like structure reaching close to the
center.

\noindent
{\it NGC~2196.} The surface brightness profile is reproduced either by
a $R^{1/4}$-law plus exponential profile with parameters as in Table
2, or by a single $R^{1/4}$-law of $\sim30''$. Visual inspection of
the image favours the first interpretation, since spiral arms are
present within 30$''$.

\noindent
{\it NGC~3054.} Regular bulge embedded in a bar-like structure typical
of early-type systems. Dust lanes across nucleus.

\noindent
{\it NGC~3067.} The central bright knot shows substructure; therefore,
it was not considered as a ``CS'' (similarly to what was done for
several galaxies of paper I, see Appendix A).

\noindent
{\it NGC~3277.} Spiral-like dust down to the nucleus.

\noindent
{\it NGC~4030.} Tightly-wound (ring-like), flocculent nuclear spiral
arms (inside the $R_e$ of the fitted exponential bulge).

\noindent
{\it NGC~4260.} The regular bulge is embedded in an early-type
bar. The profile is fitted by a $R^{1/4}$-law with $R_e=36''$ and
$V_{F606W}=11.2$. This fit clearly includes the bar component, and it
has therefore been omitted from our analysis.

\noindent
{\it NGC~4501.} Spiral-like dust lanes down to nucleus.

\noindent
{\it NGC~4806.} The profile is fitted by the sum of two exponentials.
The inner exponential has $R_e\simeq 6''$ and $V_{F606W} \simeq
14.2$. The inner component clearly includes part of the disk, and it
has been therefore excluded from our analysis. The presence of a very
small ($\lta 1''$) bulge cannot be excluded.

\noindent
{\it NGC~4980.} Faint outer disk extending to about six exponential
scales (i.e., $\approx50''$) of the fitted single-exponential
bulge. The disk shows hints for very faint spiral structure.

\noindent
{\it NGC~5121.} Nuclear elongated structure (bar?).

\noindent
{\it NGC~5377.} Tightly-wound dust lanes spiraling into nucleus, which
is surrounded by a ring-like bright structure.  The bulge-like
structure is embedded in the bar.

\noindent
{\it NGC~5448.} Strong, irregular dust lanes down to nucleus.

\noindent
{\it NGC~5985.} Nuclear spiral-like dust structure. The $R^{1/4}$-law
plus exponential fit describes the bulge and a surrounding bar.

\noindent
{\it NGC~6384.} Dust lane across nucleus. The bulge-like structure is
embedded in an early-type bar, and is fitted by the sum of two
exponentials: an inner component, reported in Table 2, and an outer
component (equivalent $R_e \sim 10''$). The two exponential components
are morphologically indistinguishable.

\noindent
{\it NGC~6810.} Central few pixels in the {\tt WFPC2} images are
saturated (central surface brightness exceeds 12.2 mag arcs$^{-2}$).

\noindent
{\it NGC~7421.} Nuclear dust lanes.  The profile is fitted by a
$R^{1/4}$-law with $R_e=20.2''$ and $V_{F606W}=12.6$. This fit clearly
includes the bar component, and it has therefore been omitted from our
analysis.

\section{Discussion}

Despite the popularity of the $R^{1/4}$-law for fitting the light
profile of the central regions of disk galaxies, growing evidence has
accumulated in the years for departures from a $R^{1/4}$ falloff of
the central light distribution, often well represented by an
exponential profile (e.g., Kormendy \& Bruzual 1978; Shaw \& Gilmore
1989; Kent, Dame \& Fazio 1991; Andredakis \& Sanders 1994). In a
recent study based on about 50 spiral galaxies, Andredakis, Peletier
\& Balcells (1995) have fitted the bulge profiles with the generalized
exponential law of S\'ersic (1968; $\Sigma(r) = \Sigma_o \exp
[-(r/r_o)^{1/n}]$, with $\Sigma(r)$ the surface brightness at the
radius $r$, $\Sigma_o$ the central surface brightness, $r_o$ a scaling
radius and $n$ the exponent variable).  These authors found that the
exponent $n$ varies with Hubble type: early spirals have essentially
$R^{1/4}$ profiles ($n=4$), and late galaxies have exponential bulges
($n=1$). Courteau, de Jong \& Broeils (1996) also found that most Sb
and later-type spirals are best fitted by a double exponential. The
{\tt WFPC2} data for our extended sample of 75 galaxies allow us to
explore the central properties of disk galaxies on scales one order of
magnitude smaller than what it is possible to probe from the ground.

Our findings are consistent and extend the general trend found from
the ground.  In Figure 5a we show the F606W bulge magnitude derived
from our fits, against the total $B$ magnitude of the
galaxy. Triangles are the bulges derived from the single-exponential
fits, pentagons those from the two-exponential fits, and the
three-vertex symbols are bulges of galaxies fitted by a single
$R^{1/4}$-law (downward) or an $R^{1/4}$-law plus exponential profile
(upward).  ESO482G17 is identified in the figure with a filled
triangle (its single-exponential fit), and by an asterisk (indicating
the upper limit to a possible bulge obtained with the
double-exponential fit); for ESO499G37, its single-exponential fit is
represented by a small filled triangle, and its large compact source
by a small asterisk (see \S 3.2). The $M (Bulge)$=$M (Galaxy)$ line is
plotted (dotted) for reference by assuming
$B-F606W=(B-F606W)_{reference}=0.65$ magnitudes, i.e. a value equal to
the estimated average $B-V$ color for the sample.  Bulge-dominated
($R^{1/4}$ or exponential) systems lie close to this relation.  Dashed
lines are plotted to indicate the $M (Bulge)$=$M (Galaxy)$ relation
for $B-F606W=(B-F606W)_{reference}\pm0.5$ magnitudes. The $B-F606W$
color of the galaxies is uncertain within several tens of a magnitude,
due to several factors. The catalog B magnitudes and the $B-V$
colors, often converted from the $B-R$ colors, are both uncertain to
$\approx 0.2$ magnitudes; the F606W filter has a different bandpass
from the $V$ filter and, particularly for red objects, the F606W
magnitudes can be significantly brighter than $V$ magnitudes
($\approx$ 0.3 magnitudes).  Exponential bulges are present in bright
($M_B(galaxy) < -19$ magnitudes), disk-dominated systems, and are
typically $\approx 2$ magnitudes fainter than their $R^{1/4}$-law
relatives for a given total galactic luminosity. In fainter hosts
($M_B(galaxy) > -19$ magnitudes), exponential bulges become the rule,
even in bulge-dominated systems.

In Figure 5b we plot the difference between the bulge magnitude
(F606W) and the total $B$ magnitude, versus Hubble type (converted
from RC3, see \S 2). Symbols are as in Figure 5a, although now thin
symbols represent isolated galaxies, thick symbols galaxies with one
or more neighbours, and large symbols galaxies which host nuclear star
formation.  ESO482G17 and ESO499G37 are excluded from this plot.  It
is clear from this figure that some level of misclassification is
present (see, e.g., the few Sb-Sc galaxies with high bulge-to-disk
ratio). Furthermore, the richness of structure observed in the inner
regions of spirals at HST resolution might suggest finer
classification schemes that the ones based on ground-based
observations.  For example, both the faint, small exponential bulges
embedded in otherwise normal outer disks (pentagons), and the systems
with an inner, dominant exponential component and an outer, quiescent
``disk'' (triangles), have morphologies dissimilar from those of
proto-typical objects of the various Hubble types; catalogs classify
the latter from very early- to very late-type spirals, although their
appearance is similar.  In spite of these caveats, the ground-based
classification generally provides an adequate {\it global} description
for most of the galaxies (as also found e.g., by Malkan, Gorjan \& Tam
1998 for their sample of active galaxies).  A possible approach is
therefore to retain such a classification, and increase the allowed
range of structural properties within each of its bins.  This avoids
further subjective arbitrariety in assigning galaxies to a specific
bin, and prevents the growth of a large number of morphological
subclasses.  In this framework, {\it (i)} ``normality'' becomes the
exception (see also Malkan et al.\ 1998); {\it (ii)} there is a
dependence of bulge profile on Hubble type (within the limitations
given by the fact that for several galaxies the surface brightness
profiles could not be obtained): while $R^{1/4}$-law bulges dominate
the earlier S0a to Sab classification bins, Sb and later spirals
preferentially host exponential bulges (in agreement with the analysis
of Courteau et al. 1996).  For Sb and later-type galaxies, the
centroid-magnitude for the exponential bulges is shifted toward
fainter magnitudes, compared to that of $R^{1/4}$ bulges.  The results
are independent of environment (see mixing of thin/thick symbols).  We
cannot exclude effects due to our limited statistics and/or selection
biases; however, at face-value, exponential bulges are typically
fainter than $R^{1/4}$ bulges, for constant total galactic luminosity
and Hubble type Sb or later.

It is remarkable that most exponential structures that we detect in
our sample host a resolved central compact source (CS). In Figure 6a
we plot the F606W magnitude of the CS against the total galactic $B$
magnitude. Triangles are the CS in single-exponential bulges;
pentagons are the CS in double-exponential galaxies; 5-vertex
asterisks are CS in galaxies with no isophotal/analytical
fit. Three-vertex symbols and 5-vertex stars are the upper limits to
the magnitudes of possible CSs in ($R^{1/4}$-law and non-fitted,
respectively) galaxies where the presence of these distinct nuclear
components is ambiguous. The enlarged sample that we have now
available puts on statistically solid grounds the relation between
central source and galaxy luminosity suggested in paper I. The
brighter the galaxy, the brighter the central source ($M_{F606W,CS}
\simeq 24.08 + 1.92 M_{B,galaxy}$ excluding the upper limits;
$M_{F606W,CS} \simeq 17.90 + 1.57 M_{B,galaxy}$ when the upper limits
are treated as measurements). The CSs in galaxies with nuclear star
formation (large symbols) tend to be brighter than those in galaxies
with quiescent nuclear appearance (small symbols; a
Kolmogorov-Smirnoff test gives a probability of 1.2\% that the two
samples are drawn from the same distribution, once the upper limits
are excluded; the probability is $< 4 \times 10^{-3}$ when the upper
limits are included as measurements). Diamonds frame those galaxies
which host a large scale or nuclear bar (solid line), or in which the
presence of a bar is suggested by e.g. an inner square-like morphology
(dotted line).  In several cases, the bar is not reported in the
catalogs classification. Many hosts of central compact sources contain
($\approx 30\%$) or are likely to contain ($\approx 30\%$) a
bar-like structure (based on the optical images).

In Figure 6b we plot the difference between the central compact source
magnitude (F606W) and the total galactic $B$ magnitude, versus
catalog Hubble type (converted from RC3). Symbols are as in Figure
6a (including large symbols for CSs in galaxies with nuclear star
formation, and small symbols for CSs in galaxies with a quiescent
nuclear appearance). In this figure, upward symbols represent isolated
galaxies, and downward symbols are galaxies with one or more
neighbours.  Central sources are present in several systems as early
as S0a-Sa, and become almost the default for intermediate type
galaxies ($\approx 60\%$ of the Sb to Sc galaxies host a central
compact source in their very center; several other S0a to Sb galaxies
have a clustering of multiple sources which has not been considered as
regular ``compact sources'' in this study, see \S 3.4 and Appendix A).
Making a clear statement about the frequency of central sources in
galaxies of different Hubble type is hampered by possible selection
effects, as demonstrated by the several upper limits which populate
the region of the diagram relative to faint compact sources in
early-type spirals. It is interesting however that the early-type disk
systems with nuclear star formation, which often makes the derivation
of the bulge light profile impractical, host the brightest central
sources (see also Appendix A).  Environment seems to have little or no
effect on this relationship: isolated and not isolated galaxies are
mixed over the entire luminosity range.

The origin of the central compact sources is unclear, and it is
possible that in some cases a contribution from an active nucleus is
present, especially for the brightest CSs in the early-type spirals.
It is in general legitimate to ask whether and how these nuclear
sources are related to the formation of the inner structures of disk
galaxies. Accurate numerical studies show that infall of dissipative
disk material, possibly forming the compact sources, might be enhanced
by the presence of a bar, and might even form (exponential) bulge-like
structures (e.g., Combes et al.\ 1990; Hasan, Pfenniger \& Norman
1993; Norman, Sellwood \& Hasan 1996, and references therein; see also
Kormendy 1993 for an extensive discussion on this issue).  The high
fraction of bars in the hosts of central compact sources and
irregular/exponential bulges might be interpreted as offering
independent support for such a scenario. Related to this, it is
legitimate to question whether the inner irregular/exponential
structures, or even some of the $R^{1/4}$-law structures, are real
``bulges'', i.e., stellar systems dynamically hotter and thicker than
the disks, or rather bars or morphologically- and/or
photometrically-distinct inner parts of the disks themselves.
Kinematic data should help us to distinguish between these scenarios.
A further discussion on the physical properties of the non-classical
inner structures is presented in Carollo (1998).

\section{Conclusions}

We have analyzed HST {\tt WFPC2} F606W images for 40 spiral galaxies.
The targets belong to the sample of 75 objects introduced in Carollo
et al. (1997a). We have studied the optical morphological properties
of the new 40 galaxies, derived the surface brightness profiles for 25
of them, and fitted the inner bulge components with an $R^{1/4}$ or
exponential law.

By using the enlarged sample of 75 galaxies, we have better quantified
the main results presented in paper I, namely: {\it I.} In about 70\%
of the galaxies there is morphological evidence for the presence of a
central component, (morphologically) distinct from the surrounding
disk. These ``bulges'' may have an irregular morphology, and
occasionally host nuclear star formation.  {\it II.}  Resolved,
central compact sources are detected in about 50\% of the
galaxies. {\it III.}  The central compact sources in (S0a-Sb) galaxies
with nuclear star formation are brighter than those hosted by galaxies
with a quiescent nuclear appearance. The luminosity of the compact
sources positively correlates with the total galactic luminosity at
the 99.9\% level.

Furthermore, the analysis of the enlarged sample of 75 objects shows
that: {\it I.}  Several of the inner, morphologically-distinct
structures are well fitted by an exponential profile.  The subset of
these ``exponential bulges'' which are embedded in otherwise normal
spiral structure appear to be fainter than $R^{1/4}$ bulges, for
constant total galactic luminosity and Hubble type. {\it II.} The
central sources are present in several systems as early as S0a-Sa;
about 60\% of the Sb to Sc galaxies host a central compact source in
their very center.  {\it III.}  The central star clusters are detected
in exponential/irregular bulges, or in bulge-less galaxies. Compact
sources embedded in $R^{1/4}$-law bulges would possibly not be
detected; however, it seems likely that if born, they would be tidally
destroyed by the high nuclear densities of these systems, see Carollo
\& Stiavelli 1998).  {\it IV.}  A large fraction ($\approx 30\%$) of
the galaxies which host a central compact source shows a barred
structure in the optical images.  In several other hosts of compact
sources ($\approx 30\%$) the presence of a bar is hinted by e.g., an
inner square-like morphology. {\it V.}  Environment seems not to play
any role in shaping the nuclear properties of disk galaxies: isolated
and not isolated galaxies show the same, and large, variety of
properties.

Our main conclusion is that in a significant fraction of galaxies
classified from the ground as relatively early-type spirals, there is
little or no morphological/photometrical evidence for a classical,
smooth, $R^{1/4}$-law bulge. Our results are fully complementary to,
and support, studies that have revealed cold kinematics in dense,
$R^{1/4}$-law structures embedded in the Sb and later-type spirals
(Kormendy 1993): both weaken the connection between ``bulges'' and low
luminosity ellipticals, and unveil a large complexity in the inner
structure of disk galaxies.

\acknowledgments

We thank Colin Norman, Tim Heckman, and the anonymous referee for
helpful comments to an earlier version of this paper.  CMC is
supported by NASA through the grant HF-1079.01-96a awarded by the
Space Telescope Institute, which is operated by the Association of
Universities for Research in Astronomy, Inc., for NASA under contract
NAS 5-26555.  MS and JM acknowledge support from DDRF grant
D001.82148. MS acknowledges support from the Italian Space Agency.
This research has been partially funded by GRANT GO-06359.01-95A
awarded by STScI, and has made use of the NASA/IPAC Extragalactic
Database (NED) which is operated by the Jet Propulsion Laboratory,
Caltech, under contract with NASA.

\bigskip
\bigskip

\noindent
{\bf Appendix A}

In order to homogenize the measurements for the enlarged sample of 75
galaxies, we performed the exponential fits described in \S 3.2 also
for the galaxies presented in paper I.  In addition, in contrast with
paper I and consistently with \S 3.2, we excluded from the fits the
central compact sources, and repeated the $R^{1/4}$ fits for those
galaxies hosting such a component.  We adopted as final the best of
the two fits ($R^{1/4}$-law or exponential). As a result, several
galaxies for which in paper I the analytical fits had identifyed an
$R^{1/4}$-law component are now listed instead as hosts of exponential
bulges. The new fits have a better $\chi^2$, and generally provide a
better description of the morphologically-distinct structures seen in
the images.

The bulge and central compact source parameters for the galaxies of
paper I used in the present analysis are given in Table 3.  For easy
referencing, we identify in the Table the source for the analytical
fits to the bulges. We have included in the list also those galaxies
which have no good fit or no surface brightness profile, but host a
central compact source [whose parameters are taken from paper I, and
have been further checked with method (3) of \S 3.3]. Six galaxies
reported as hosts of central compact sources in paper I, namely
NGC2397, NGC2964, NGC3885, NGC5879, NGC6000 and IC4390, have been
excluded from the present analysis, since in these objects the central
source shows substructure and/or might be affected by dust
obscuration. Four of the six excluded compact sources are hosted by
galaxies with nuclear star formation and morphological types in the
range S0a--Sb. They are among the brightest in the sample, and if
included, would strengthen the finding that the brightest sources are
embedded in the early- to intermediate-type spirals.

\newpage
\setcounter{page}{29}
\begin{figure}
\caption{Inner {\tt WFPC2} F606W $18'' \times 18''$ of the 40
galaxies. North is up, East is left. These images would only have a
few resolution elements in the Palomar Observatory Sky Survey, and the
structure we detect would disappear.}
\end{figure}

\begin{figure}
\caption{Unsharp-masking images ($9'' \times 9''$) for NGC 3177, NGC
5121, NGC 5377 and NGC 7690. North is up, East is left. }
\end{figure}

\begin{figure}
\caption{Surface brightness profiles for 25 galaxies of the sample.
Shown are the analytical fits to the inner components: solid lines for
$R^{1/4}$-law fits, dashed lines for exponential fits. Several
galaxies host an exponential bulge.}
\end{figure}

\begin{figure}
\caption{The {\tt WFPC2} mosaiced images for four galaxies
representative of the two classes of hosts of exponential bulges. The
upper panels show two ``bulge-dominated'' objects fitted by
single-exponential profiles (ESO508G34, left; NGC4980, right). The
lower panels show two ``disk-dominated'', double-exponential objects
(NGC4030, left; NGC6384, right). The arrows and the orthogonal
segments indicate the North and East directions, respectively. They
are  $\approx 12''$ long.}
\end{figure}

\begin{figure}
\caption{Panel a: The F606W bulge magnitude versus total galactic $B$
magnitude. Triangles represent the bulge-dominated galaxies fitted by
a single exponential profile; pentagons are the bulges of
disk-dominated galaxies fitted by a double-exponential profile; the
three-vertex symbols are bulges of galaxies fitted by a single
$R^{1/4}$-law (downward) or an $R^{1/4}$-law plus exponential profile
(upward).  The single-exponential fit for ESO482G17 is represented by
the (large) filled triangle; the large asterisk represent the upper
limit obtained with a double-exponential fit to a possible bulge
component in this galaxy. The single-exponential fit for ESO499G37 is
represented by the small filled triangle; its compact source, the
largest of our sample, is represented by the small asterisk.  The
$M(Bulge)$=$M(Galaxy)$ line is plotted (dashed) by assuming
$B-V=0.65$, i.e. the estimated average $B-V$ color for the sample.
Dashed lines indicate the $M(Bulge)$=$M(Galaxy)$ relation for
$B-F606W=0.65\pm0.5$ magnitudes.  Panel b: Bulge minus galaxy
``delta-magnitude'', i.e. F606W bulge magnitude minus total galactic
$B$ magnitude, versus morphological type (obtained converting the RC3
classification, see text). Dashed line and symbols are as in panel
a. Thin symbols represent isolated, and thick symbols non-isolated
galaxies, respectively; large symbols represent galaxies which host
nuclear star fomation. ESO482G17 is excluded from this panel. }
\end{figure}

\begin{figure}
\caption{Panel a: The F606W magnitude of the central compact source
versus the total galactic $B$ magnitude. Triangles are the CSs in
galaxies with a single-exponential bulge; pentagons are those embedded
in double-exponential galaxies; five-vertex asterisks are the CSs in
galaxies with no isophotal/analytical fit.  Three-vertex symbols and
5-vertex stars are the upper limits to the magnitudes of possible CSs
in ($R^{1/4}$-law and non-fitted, respectively) galaxies where the
presence of these distinct nuclear components is ambiguous.  Large
symbols are assigned to galaxies with nuclear star formation; small
symbols represent galaxies with a quiescent nuclear appearance.
Diamonds frame the symbols relative to galaxies which host (solid
line) or are likely to host (dotted line) a bar. Panel b: Central
compact source minus galaxy ``delta-magnitude'', i.e. F606W magnitude
of the central compact source minus total galactic $B$ magnitude,
versus Hubble type (converted from RC3). Symbols as in panel a.}
\end{figure}

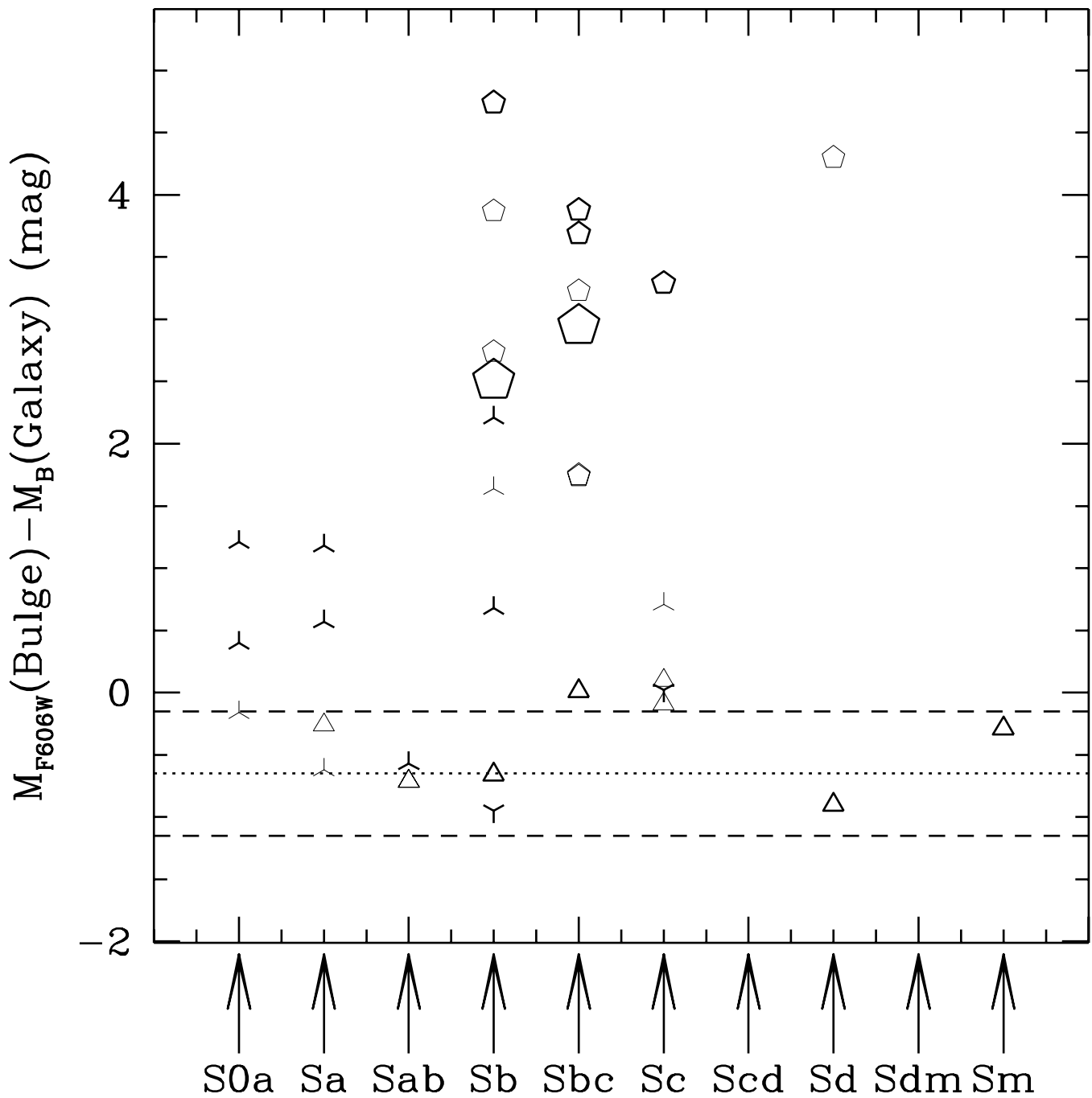
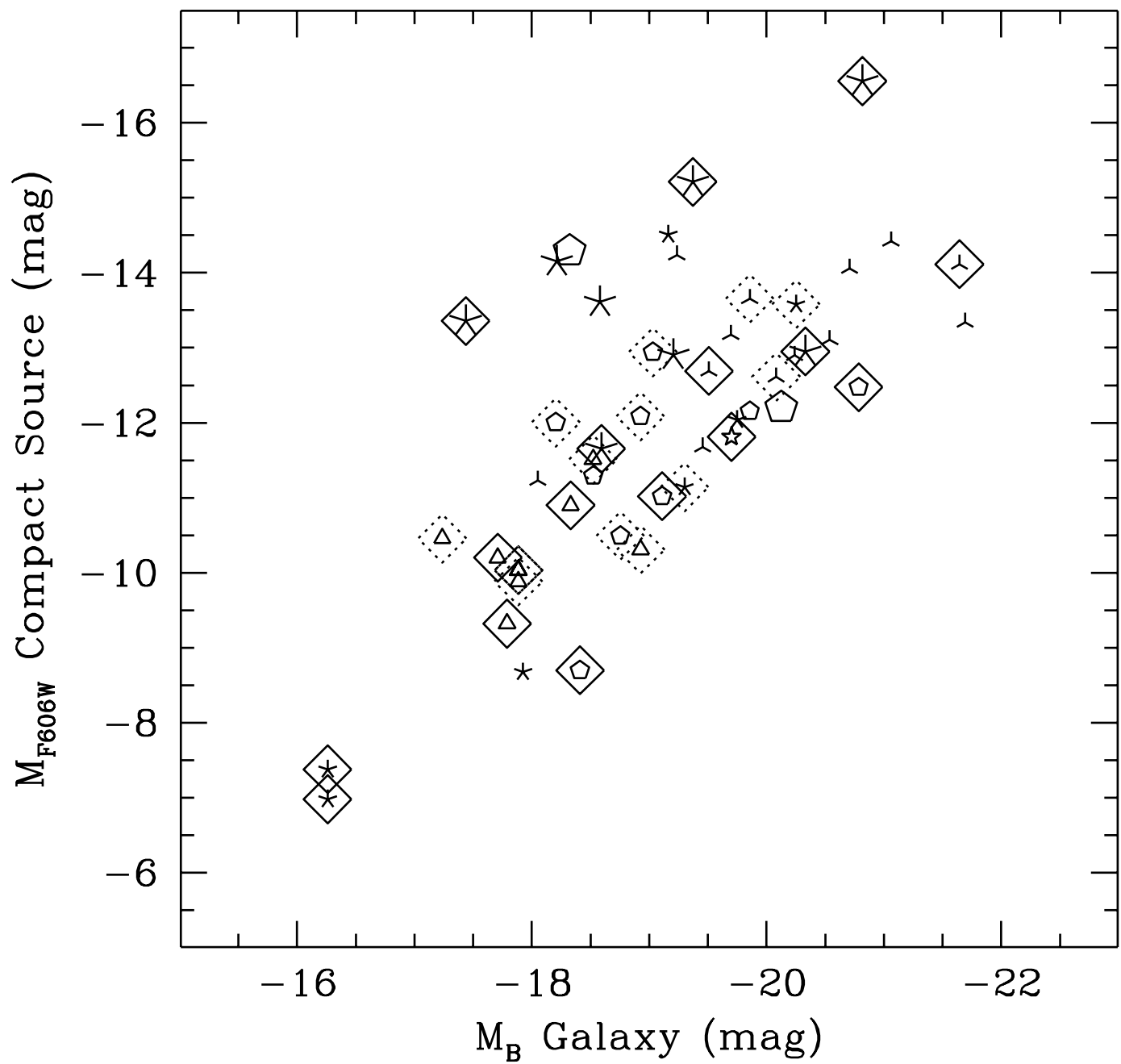
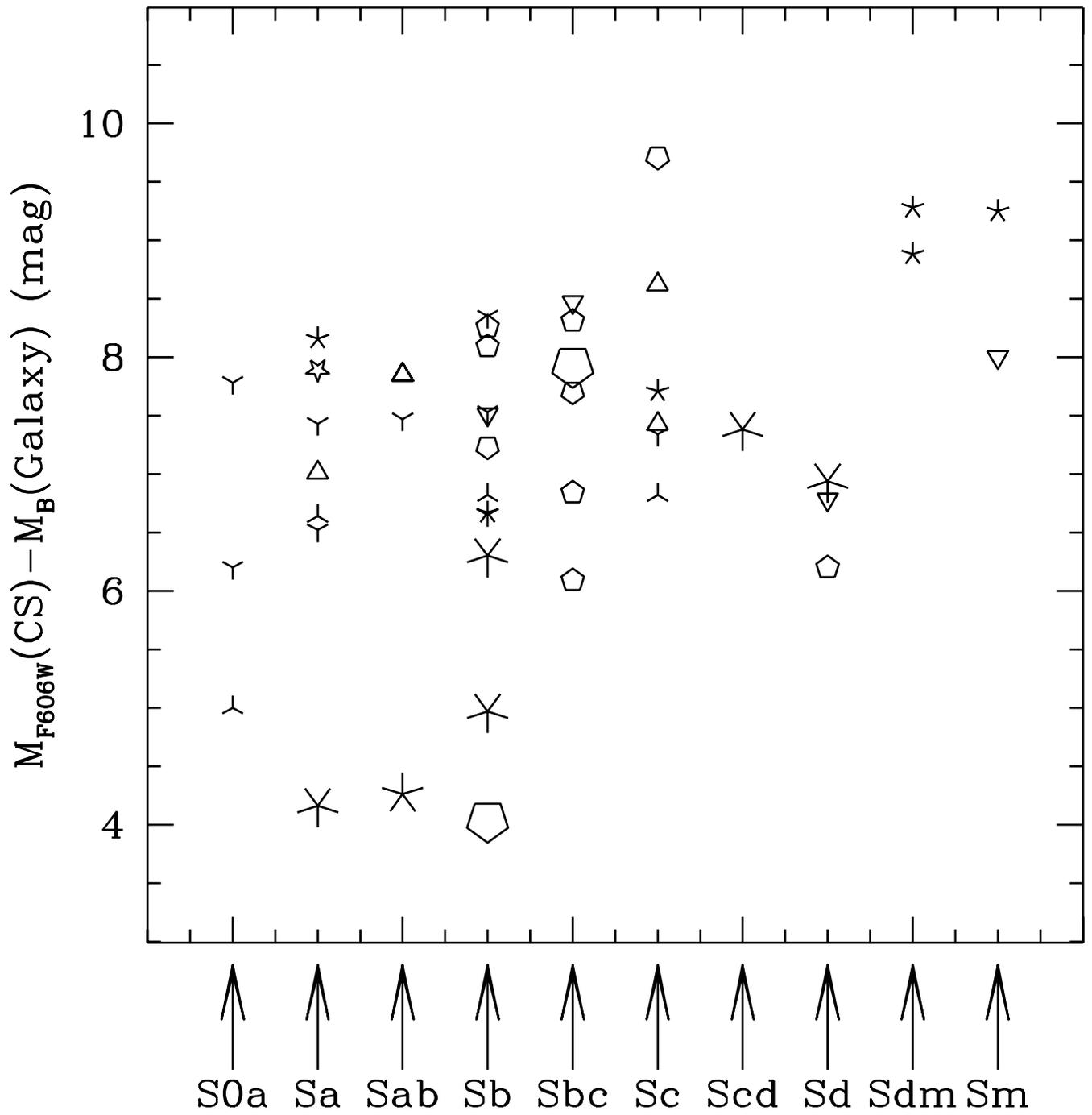
\end{document}